# Further studies of GEM performance in dense noble gases

V. Aulchenko, A. Bondar, A. Buzulutskov[*], L. Shekhtman, R. Snopkov, Yu. Tikhonov

*Budker Institute of Nuclear Physics, pr. Lavrentieva 11, 630090 Novosibirsk, Russia*

**Abstract**

We further study the performance of single, double and triple Gas Electron Multiplier (GEM) detectors in pure noble gases at high pressures, in the range of 1-10 atm. We confirm that light noble gases, in particular He and its mixtures with Kr, have the highest gain, reaching $10^6$, and an unusual gain dependence on pressure. Effects of the number of GEMs, GEM hole diameter and pitch are investigated in detail. In He, avalanche-induced secondary scintillations are observed at high gains, using the metal photocathode. These results are relevant in the field of avalanche mechanism in noble gases and X-ray, neutron and cryogenic particle detectors. © 2003 Elsevier Science. All rights reserved

## 1. Introduction

Operation of Gas Electron Multiplier (GEM) [1] detectors in dense noble gases, i.e. at high pressures or at low temperatures, is relevant in the field of cryogenic particle detectors [2], neutron [3] and X-ray imaging [4] and gaseous photomultipliers [5]. It turned out that light noble gases, He and Ne, had superior performances at high pressures in terms of the maximum gain and stability of operation [6,7]. In addition, He and Ne showed an unusual gain dependence on pressure [7]: in a wide pressure range the operation voltage does not increase with pressure. Using a simple model of avalanche development, this was interpreted as a result of a new avalanche mechanism arising in He and Ne at high gas densities, namely the associative ionization [2].

In this paper we further study the performance of different GEM structures in noble gases, at high pressures, and in particular in He, Kr and their mixture. He and Kr can be considered as typical examples of light and heavy noble gases, since the GEM operation in these gases is rather similar to that of Ne and Xe, respectively [7]. The effects of the number of GEMs in cascade and GEM hole diameter and pitch are studied. Photon feedback effects in He are also investigated.

## 2. Results

One, two or three GEM foils, a Cu cathode and a printed-circuit-board (PCB) anode were mounted in cascade inside a high-pressure vessel. The GEMs of three types were used with the following hole diameter/pitch values: 80/140, 40/140 and 40/100μm. The vessel was filled with pure He, Kr or their mixture. The detector was irradiated with an X-ray tube. In triple, double and single GEM detectors the anode signal was read out from the PCB. These modes of operation are designated as 3GEM+PCB, 2GEM+PCB and 1GEM+PCB, respectively. The gain values were determined in the current mode of operation. More detailed description of the experimental setup and procedures are presented elsewhere [6,7].

---

[*] Corresponding author. Tel.: +7-3832-394833; fax: +7-3832-342163; e-mail: buzulu@inp.nsk.su.

Anode signals of a triple GEM detector in He and Kr are shown in Fig. 1, after a charge-sensitive amplifier. In Kr at all gains and in He at gains below few thousands (not shown) the anode pulse has a triangular shape: a linear pulse rise corresponds to the integration of the primary signal induced by electron avalanching. The width of the primary signal (FWHM) is about 150 and 100 ns in He and Kr, respectively. At high gains, exceeding $10^4$, the primary signal in He is accompanied by a secondary signal (Fig.1, top and middle). The latter is due to photon feedback between the last GEM element and Cu cathode, induced by secondary scintillations in He under avalanche conditions.

Indeed, similarly to other noble gases He is a good scintillator in the vacuum ultraviolet (VUV) region, with an emission centered at 80 nm [8]. And the quantum efficiency of Cu photocathode at this wavelength is high enough, of about 10% [9], to provide the efficient photon feedback. On the other hand, the quantum efficiency of Cu is considerably reduced at the emission region of Kr (150 nm). Therefore the anode signal in Kr has no secondary signal (Fig.1, bottom).

In addition, the photon feedback in the double GEM is stronger than that of the triple GEM (Fig.1): at a gain of $10^4$ it amounts to 100% and 20% of the primary signal, respectively. This is because of the screening effect of the multi-GEM structure: the optical transparency of each GEM is only 10%. The characteristic time of He emission, in a gas discharge, was reported to be of the order of 1 μs [8], which is compatible with our observations (Fig.1).

Fig.2 shows gain-voltage characteristics of a triple GEM detector in He and He+1%Kr, at different pressures. In these measurements, the photon feedback was reduced compared to that of Fig.1 and ref. [7]: the middle GEM had a reduced hole diameter (40 μm). Therefore somewhat higher gains, reaching $10^6$, are obtained here, compared to the previous results [7].

One can see, that in the range of 1-5 atm the operation voltage in He almost does not change, in accordance with earlier observation [7]. In [2] this behavior was explained by a new avalanche mechanism, namely the associative ionization, taking over the impact ionization at high pressures. In the associative ionization, the electron is produced via association of an atom with an excited atom into a molecular ion [10]: $He + He^* \rightarrow He^+_2 + e^-$.

As was pointed out in [2], at the background of the associative ionization the effects of impurities and in particular the Penning effect are not seen at high pressures. This statement is supported by gain-voltage characteristics in He+1%Kr (Fig.2), which is a Penning mixture by definition. Indeed, it is well known that at low pressures Penning effect results in lowering the operation voltage, compared to pure gas. However, this is not the case at high pressures: one can see that the operation voltage in He+Kr either does not differ from that of pure He (at 1atm) or increases with pressure.

It is interesting that at gains exceeding $10^5$, the gain increase with voltage is faster than exponential in He and slower than exponential in He+Kr. The former is obviously due to secondary scintillations in He discussed above. The latter means that the scintillation of He specie is quenched, in presence of Kr, and that the avalanche is saturated. The avalanche saturation at these relatively low gains might be due to the effect of avalanche confinement within the GEM hole considered in [11].

Optimization of GEM hole diameter/pitch values for high pressure operation was one of the goal of the current study. Fig.3 shows gain-voltage characteristics of a single GEM detector in He and Kr, at 5 atm, for three different configurations: 80/140, 40/140 and 40/100 μm. One can see that decreasing GEM hole diameter does not improve the operation at high pressures. In terms of the maximum gain, the best configuration turned out to be the "standard" one: 80/140 μm.

The other goal of the current work was to reach as high as possible operation pressures in Kr and Xe, which is relevant in the field of X-ray detection. So far, it did not exceed 5 atm in triple-GEM detectors [7], because the voltage applied across each GEM could not exceed a certain value, namely 700 V in heavy noble gases. This was explained by ion feedback between GEM elements, resulting in ion-induced electron emission from the GEM electrode [2]. Such an emission is particular enhanced in noble gases [10]. Thus, suppressing the ion feedback, for example by operating with a single GEM detector rather than with that of multi-GEM, one might expect reaching higher operation pressures. This turned out to be the case as one can see from Fig.4: it shows gain-voltage characteristics of a single GEM detector in Kr at different pressures. Indeed, here the maximum operation pressure and voltage exceed 10 atm and 1200 V, respectively, the maximum gain having a moderate value, of the order of $10^2$, even at 10 atm.

## 3. Conclusions

In conclusion, we further studied the performance of single, double and triple GEM detectors in pure He and Kr at high pressures, in the range of 1-10 atm.

We confirm that light noble gases, in particular He and its mixtures with Kr, have the highest gain, reaching $10^6$. This was explained by the hypothesis that the associative ionization is the dominant avalanche mechanism in dense He and Ne [2]. It has however a minor effect in heavy noble gases. The practical consequence is that higher gains than expected are reached in dense He and Ne. This property is very attractive for applications in cryogenic two-phase particle detectors.

In He, avalanche-induced secondary scintillations are observed at high gains, using the Cu photocathode. This observation might open the way for developing He-based gas scintillation detectors, using a large variety of efficient photocathodes stable in air (metals, oxides, etc). The apparent application of He-based detectors in general is a non-ageing neutron detector in compressed $He^3$.

We also studied the effect of the GEM hole diameter and pitch and the number of GEMs on the performance at high pressures. It is possible to have moderate gains (~100) at rather high pressures (10 atm) in heavy noble gases if to do without ion feedback between GEM elements, namely to operate with a single GEM detector, with a larger hole diameter. This result may be attractive for X-ray detectors in compressed Xe and Kr.

Further studies of this technique, i.e. GEM operation at cryogenic temperatures, in two-phase detectors, at higher pressures, in other non-ageing gases, are on the way.

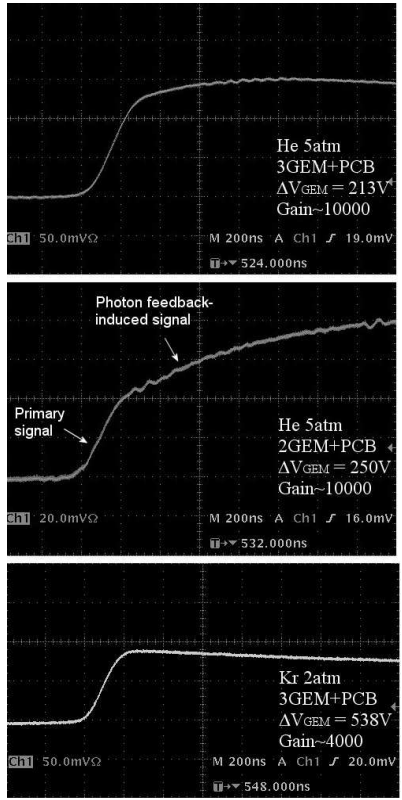

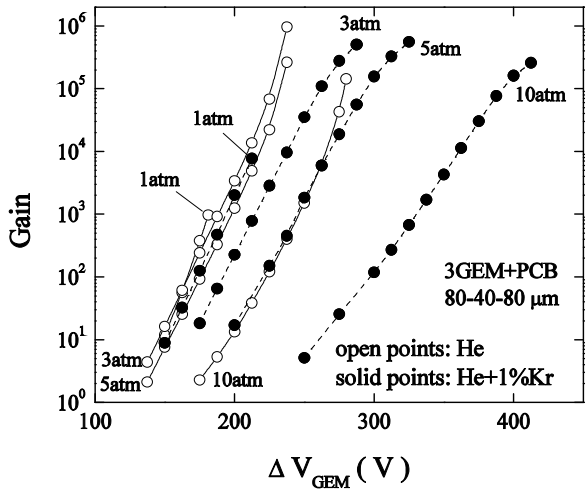

Fig.1 Anode signals in He at 5 atm, in 3GEM+PCB (top) and 2GEM+PCB (middle) detectors at a gain of 10000, and in Kr at 2 atm in 3GEM+PCB detector at a gain of 4000. All GEMs have 80/140 μm hole diameter/pitch configuration.

Fig.2 Gain of a 3GEM+PCB detector at different pressures in He and He+1%Kr as a function of the voltage across each GEM. All GEMs have 80/140 μm hole diameter/pitch configuration, except of the middle GEM which has a reduced hole diameter, of 40 μm.

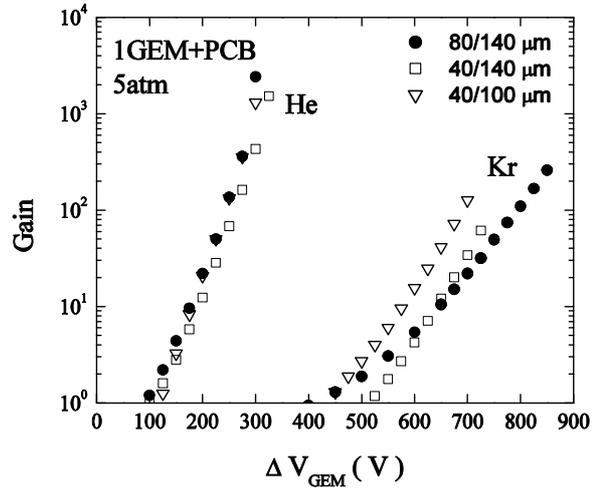

Fig.3 Gain-voltage characteristics of a 1GEM+PCB detector in He and Kr, at 5 atm, for three different GEM hole/pitch configurations: 80/140, 40/140 and 40/100 μm.

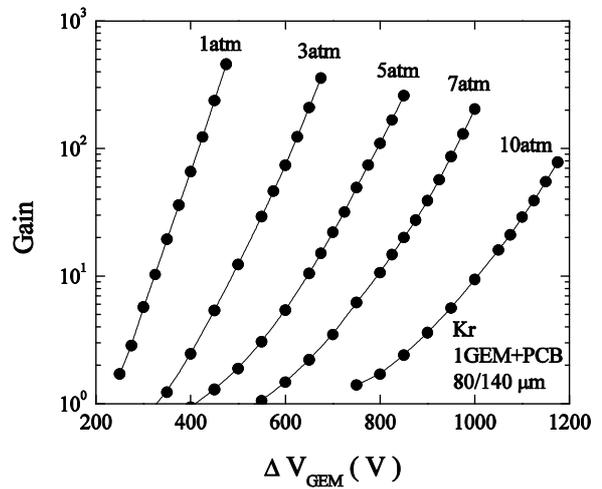

Fig.4 Gain-voltage characteristics of a 1GEM+PCB detector in Kr at different pressures. GEM hole/pitch configuration is 80/140 μm.